\documentstyle[twoside,fleqn,espcrc2,epsf]{article}

\newcommand{\half}{\mbox{\small $\frac{1}{2}$}}          
\newcommand{\third}{\mbox{\small $\frac{1}{3}$}}         
\newcommand{\msbar}{\mbox{\tiny $\overline{MS}$}}        
\newcommand{\rgi}{\mbox{\tiny $RGI$}}                    

\def\lsim{\mathrel{\rlap{\lower4pt\hbox{\hskip1pt$\sim$}}
    \raise1pt\hbox{$<$}}}                
\def\gsim{\mathrel{\rlap{\lower4pt\hbox{\hskip1pt$\sim$}}
    \raise1pt\hbox{$>$}}}                

\def\3{\ss}

\newcommand{\AmS}{{\protect\the\textfont2
  A\kern-.1667em\lower.5ex\hbox{M}\kern-.125emS}}

\title{
       \vspace{-3.65cm}                                     %
       {\normalsize DESY 99--114}    \\[-0.2cm]             
       {\normalsize HUB--EP--99/36}  \\[-0.2cm]             
       {\normalsize FUB-HEP/4-99}    \\[-0.2cm]             
       {\normalsize HLRZ 99-35}      \\[-0.2cm]             
       {\normalsize August 1999}     \\                     
       \vspace{1.32cm}                                      
       Light quark masses from the lattice%
            \thanks{Talk given by R. Horsley at Lat99,      
                    Pisa, Italy.}}                          

\author{M.~G\"ockeler%
           \address{Institut f\"ur Theoretische Physik, Universit\"at
                    Regensburg, D-93040 Regensburg, Germany},
        R.~Horsley%
           \address{Institut f\"ur Physik, Humboldt-Universit\"at zu Berlin,
                    D-10115 Berlin, Germany},
        B.~Klaus%
           \address{Institut f\"ur Theoretische Physik,
                    Freie Universit\"at Berlin, D-14195 Berlin, Germany},
        W. K\"urzinger$^{\rm c,}$%
           \address{Deutsches Elektronen-Synchrotron DESY \& NIC,
                    D-15735 Zeuthen, Germany},
        H.~Oelrich$^{\rm d}$,
        D.~Petters$^{\rm c,}$ \hspace{-0.25cm} $^{\rm d}$,
        D.~Pleiter$^{\rm c,}$ \hspace{-0.25cm} $^{\rm d}$,
        P.~E.~L. Rakow$^{\rm a}$,
        G.~Schierholz$^{\rm d,}$%
           \address{Deutsches Elektronen-Synchrotron DESY,
                    D-22603 Hamburg, Germany}
        and
        P.~Stephenson%
           \address{Dipartimento di Fisica,
                    Universit\`a degli Studi di Pisa \& INFN,
                    Sezione di Pisa, 56100 Pisa, Italy}}
       
\begin{document}

\begin{abstract}
A completely non-perturbative estimate is given for the $u/d$
and strange quark masses in quenched QCD using $O(a)$
improved fermions and, for comparison, Wilson fermions.
For improved fermions we find
$m_{u/d}^{\msbar}(\mu=2\,\mbox{GeV}) = 4.4(2)\,\mbox{MeV}$,
$m_s^{\msbar}(\mu=2\,\mbox{GeV}) = 105(4)\,\mbox{MeV}$
when using $r_0$ to set the physical scale.
\end{abstract}

\maketitle

\setcounter{footnote}{0}


\section{THE LATTICE APPROACH}

Lattice methods allow, in principle, an `ab initio'
calculation of the fundamental parameters of QCD, among them
the quark masses. These, not being asymptotic states
of the Lagrangian, need to be defined
by giving the scheme ${\cal S}$ and scale $M$.
In this brief report we give our recent results for the light quark
masses. Further details may be found in \cite{gockeler99a}.

The starting point is the PCAC quark mass $a\widetilde{m}_q$,
which can be determined from the ratio of the correlation functions
\begin{eqnarray}
   a\widetilde{m}_{q_1} + a\widetilde{m}_{q_2}
    \stackrel{t\gg 0}{=}
       { \langle
            \partial_4 {\cal A}_4^{q_1q_2}(t){\cal P}^{q_1q_2}(0)
          \rangle \over
            \langle {\cal P}^{q_1q_2}(t){\cal P}^{q_1q_2}(0) \rangle },
                                                        \nonumber
\end{eqnarray}
where ${\cal A}$ and ${\cal P}$ are the axial and pseudoscalar currents,
respectively, for possibly non-degenerate quarks $q_1$ and $q_2$.
Expanding $a\widetilde{m}_{q_1}+a\widetilde{m}_{q_2}$
(with expansion coefficients $\widetilde{Y}$, $\widetilde{c}$) and the 
pseudoscalar mass $am_{PS}^{q_1q_2}$ (with expansion
coefficients $Y_{PS}$, $c_{PS}$) in terms
of the masses $am_{q_i}$
($\equiv \half \left(1/\kappa_{q_i} - 1/\kappa_c\right)$, $i= u/d, s$)
and renormalising to the `renormalisation group invariant'
(RGI) form gives
\begin{eqnarray}
   r_0 m^{\rgi}_{u/d} &=& c^*_a (r_0 m_{\pi^+})^2 +
                            c^*_b (r_0 m_{\pi^+})^4,
                                             \nonumber
\end{eqnarray}
and
\begin{eqnarray}
   \lefteqn{r_0 m^{\rgi}_s =}
      &                                    \nonumber \\
      & c^*_a \left[ (r_0 m_{K^+})^2 + (r_0 m_{K^0})^2
                                    - (r_0 m_{\pi^+})^2 \right] + &
                                           \nonumber \\
      & c^*_b \left[ (r_0 m_{K^+})^4 + (r_0 m_{K^0})^4
                                    - (r_0 m_{\pi^+})^4 \right]. &
                                           \nonumber
\end{eqnarray}
We have defined $m^{\rgi}_{u/d} = (m^{\rgi}_u + m^{\rgi}_d)/2$,
and have ignored any
small corrections due to electromagnetic effects.
The `force' scale \cite{guagnelli98a} has been used to set the
physical scale and the $c^* \equiv \lim_{g_0\to 0} c $
coefficients are given by
\begin{eqnarray}
     c_a &=& F \left[{\widetilde{Y}\over Y_{PS}}\right]
                       \left({r_0\over a}\right)^{-1},
                                                        \nonumber \\
     c_b &=& F \left[{\widetilde{Y}\over Y_{PS}}\right]
                      \left[{-c_{PS}\over Y_{PS}}
                         \left( {r_0\over a} \right)^{-2} \right]
                       \left({r_0\over a}\right)^{-1},
                                                        \nonumber
\end{eqnarray}
where $F \equiv \Delta Z^{\cal S}(M) \widetilde{Z}_m^{\cal S}(M)$
with $\widetilde{Z}_m^{\cal S}(M)$ the renormalisation constant
and $\Delta Z^{\cal S}(M)$ converting the
renormalised mass to the {\it RGI} mass $m^{\rgi}$.
Our convention is to use 
\begin{eqnarray}
   \lefteqn{[\Delta Z_m^{\cal S}(M)]^{-1} =}
      &  &                                              \nonumber \\
      &  & \left[ 2b_0 g^{\cal S}(M)^2 \right]^{d_{m0}\over 2b_0}
           e^{\int_0^{g^{\cal S}(M)} d\xi
           \left[ {\gamma_m^{\cal S}(\xi)
                             \over \beta^{\cal S}(\xi)} +
                 {d_{m0}\over b_0 \xi} \right] }.
                                                        \nonumber
\end{eqnarray}
(In the $\overline{MS}$ scheme $\beta^{\msbar}$ and $\gamma^{\msbar}_m$
are known to four loops.)

\begin{figure}[htb]
   \vspace*{-0.25in}
   \epsfxsize=7.00cm \epsfbox{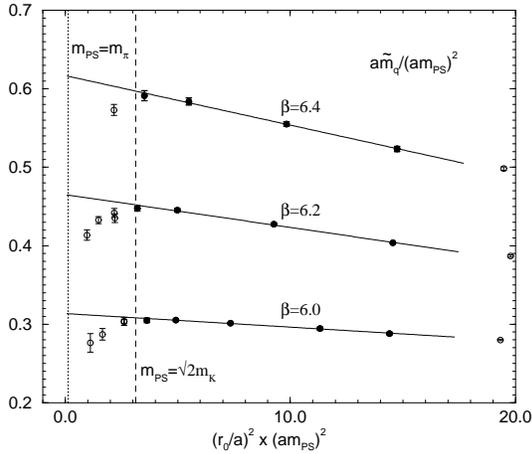}
   \vspace*{-0.25in}
   \caption{\footnotesize{\it The ratio $a\widetilde{m}_q/(am_{PS})^2$
            against $(r_0/a)^2 \times (am_{PS})^2$ for $\beta =$
            $6.0$, $6.2$, $6.4$. The filled circles are the
            data used in the fits. The dashed line ($\sim 3.13$)
            corresponds to $m_{PS}=\sqrt{2}m_K$, while the dotted line
            ($\sim 0.125$) is $m_\pi$.}}
   \vspace*{-0.25in}
   \label{fig_mqwiompi2_mpi2_magic_lat99}
\end{figure}

To determine the expansion coefficients
we shall consider degenerate quarks,
\begin{eqnarray}
   a\widetilde{m}_q &=& \widetilde{Y}
                    \left[ 1 + (\widetilde{c}+d)am_q \right] am_q,
                                                        \nonumber  \\
   (am_{PS})^2  &=& Y_{PS} \left[ 1 + (c_{PS}+d)am_q \right] am_q,
                                                        \nonumber
\end{eqnarray}
giving
\begin{eqnarray}
   { a\widetilde{m}_q \over (am_{PS})^2 } = {\widetilde{Y}\over Y_{PS}}
                \left[ 1 + \left( { \widetilde{c} - c_{PS} \over Y_{PS} }
                           \right) (am_{PS})^2
                \right].
                                                        \nonumber
\end{eqnarray}
While the constant term on the r.h.s. of
this equation is sufficient to find $c_a$,
the gradient term does not fully determine $c_b$.
(We shall assume in the following that $\widetilde{c}$
is small \cite{gockeler99a}.)
In Fig.~\ref{fig_mqwiompi2_mpi2_magic_lat99} we show this ratio
for $O(a)$ improved fermions. 
Below $m_q \sim m_s$ there are significant deviations
from linearity, due to the presence of chiral logarithms.
For large quark masses, we expect non-linear terms to appear.
From the figure a safe linear region would seem to be
$m_s \lsim m_q \lsim \third m_c \sim 3m_s$
(we have $2(r_0m_D)^2 \sim 44.9$).

Similar results also hold for Wilson fermions
(although there we only have two $\beta$ ($\equiv 6/g_0^2$) values,
namely $6.0$ and $6.2$).

We also need $F(g_0)$.
For $O(a)$ improved fermions this was recently achieved
by the ALPHA collaboration \cite{capitani98a},
using the Schr\"odinger Functional formalism, giving
$F(g_0)$ for $6.0 \le \beta \le 6.5$.
For Wilson fermions \cite{gockeler98a} we used the method proposed
in \cite{martinelli94a}, which mimics perturbation theory in a {\it MOM}
scheme by considering amputated quark Green's function,
with an operator insertion. By `converting' the {\it MOM}
scheme to the $\overline{MS}$ scheme \cite{franco98a},
we have found $F(g_0)$ at $\beta = 6.0$, $6.2$.


\section{CONTINUUM RESULTS}

\begin{figure}[htb]
   \vspace*{-0.25in}
   \epsfxsize=7.00cm \epsfbox{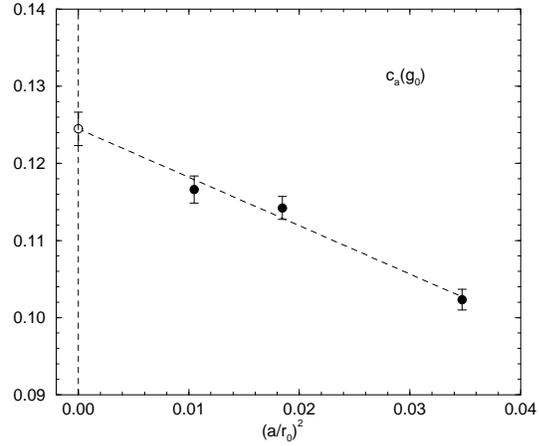}
   \vspace*{-0.25in}
   \caption{\footnotesize{\it The continuum extrapolation for
            $c_a$ for $O(a)$ improved fermions.}}
   \label{fig_ca_rgi_magic_lat99}
   \vspace*{-0.25in}
\end{figure}

We now have all the pieces necessary to compute $c_a$ (and $c_b$
approximately) and to extrapolate to the continuum limit.
In Fig.~\ref{fig_ca_rgi_magic_lat99}
we plot our results for $c_a$ for $O(a)$ improved fermions.
Similar results also hold for $c_b$ and for Wilson fermions.
(The latter are, of course, extrapolated in $a$ rather than $a^2$.)
Using these results and the physical kaon and pion masses, we find
\begin{eqnarray}
   m_{u/d}^{\rgi} = 6.1(2) \, \mbox{MeV}, \quad
   m_s^{\rgi} = 146(4) \, \mbox{MeV}.
                                           \nonumber
\end{eqnarray}
(Note that we have used $r_0 = 0.5\,\mbox{fm}$ to set the scale.
Other choices can lead to an $O(10\%)$ difference.)
For Wilson fermions we have
$m_{u/d}^{\rgi} = 5.3(8)\,\mbox{MeV}$,
$m_s^{\rgi} = 121(20)\,\mbox{MeV}$.
These results are somewhat lower than the $O(a)$ improved
numbers but with far larger error bars.
Note that as we only have two values of
$\beta$, this makes a continuum extrapolation more difficult.
Also the number of $\kappa$ values
used and the size of the data sets are smaller than
for $O(a)$ improved fermions. Nevertheless, within a one-standard
deviation the results are in agreement.

In the $\overline{MS}$ scheme at the `standard' value
of $\mu = 2\,\mbox{GeV}$ we find for $O(a)$ improved fermions
\begin{eqnarray}
    m^{\msbar}_{u/d} = 4.4(2) \,\mbox{MeV}, \quad
    m^{\msbar}_s     = 105(4) \,\mbox{MeV}.
                                           \nonumber
\end{eqnarray}
The equivalent Wilson results are $3.8(6)\,\mbox{MeV}$ and
$87(15)\,\mbox{MeV}$ for the $u/d$ and strange quark masses respectively.

For the $m_{u/d}$ quark mass result we have simply
extrapolated the fits for the strange quark mass result.
The mass ratio $m_s/m_{u/d}$ for $O(a)$ improved fermions
is $\approx 23.9$, which is very close to the value given in
leading order chiral perturbation theory, namely
$(m_{K^+}^2 + m_{K^0}^2-m_{\pi^+}^2) / m_{\pi^+}^2 \approx 24.2$.
This is simply because $|c_b^*| \ll |c_a^*|/(r_0m_K)^2$, and so
the second term in the Taylor series expansion is negligible.
The mass ratio is then independent of $c_a^*$.


\section{COMPARISON WITH OTHER RECENT RESULTS}

\begin{figure}[htb]
   \vspace*{0.15in}
   \epsfxsize=7.00cm \epsfbox{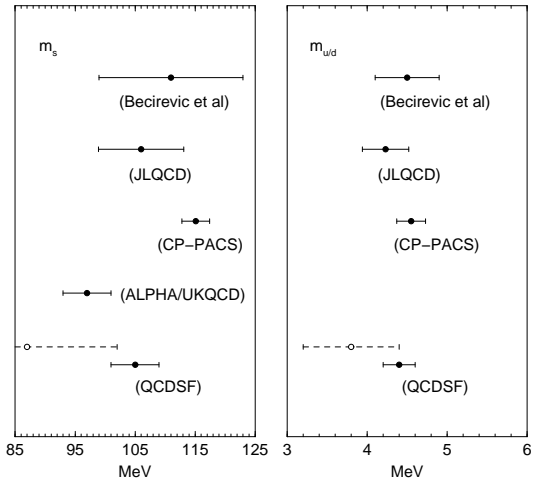}
   \vspace*{-0.25in}
   \caption{\footnotesize{\it Some recent quark mass results
            in the $\overline{MS}$ scheme at a scale of $\mu = 2\,\mbox{GeV}$.
            The references are given in the text. Our Wilson
            quark mass determination is denoted by dashed
            error bars.}}
   \label{fig_qm_comparison_lat99}
   \vspace*{-0.25in}
\end{figure}

We shall now briefly compare our results (QCDSF) with other recent
quark mass determinations (published since the last lattice
conference). In Fig.~\ref{fig_qm_comparison_lat99}
we show the results from \cite{becirevic98a,jlqcd99a,cppacs99a} and
\cite{garden99a}. Closest to the method used here is \cite{garden99a}
(ALPHA/UKQCD). We have not converted the different physical scales
used in \cite{becirevic98a,jlqcd99a,cppacs99a}
to the $r_0$ scale. The quark masses are determined from
the pion and kaon. Ref.~\cite{becirevic98a} uses $O(a)$ improved
fermions mainly at $\beta=6.2$ and the method of
\cite{martinelli94a} for the renormalisation constants,
while \cite{jlqcd99a} (JLQCD) uses staggered Wilson fermions
at three $\beta$ values and also \cite{martinelli94a} for the
renormalisation constants. Ref.~\cite{cppacs99a} (CP-PACS)
uses Wilson fermions (with four $\beta$ values and a one-loop
perturbative renormalistion factor).
Within a rough $10\%$ band there is agreement.


\section*{ACKNOWLEDGEMENTS}

The numerical calculations were performed on the
Quadrics {\it QH2} at DESY (Zeut\-hen) as well as
the Cray {\it T3E} at ZIB (Berlin) and the Cray {\it T3E} at
NIC (J\"ulich).



\end{document}